\documentclass[aps,prb,superscriptaddress,papersize=a4paper,twocolumn,10pt,showpacs]{revtex4}
\usepackage{amsmath,amsfonts,amssymb,amsthm}
\usepackage{mathtext,mathtools,mathrsfs, bm, comment}
\usepackage{esint,bm,esvect,dsfont}
\usepackage{graphicx,epsfig,psfrag,xcolor}
\usepackage{array,tabularx,tabulary,booktabs,multirow}
\graphicspath{{figures/}}
\usepackage{hyperref}
\usepackage{verbatim}
\usepackage{float}

\def\ve{\varepsilon}

\def\vf{\varphi}
\def\de{\partial}

\def\bq{\mathbf{q}}
\def\bA{\mathbf{A}}

\def\de{\partial}

\def\ve{\varepsilon}

\def\vf{\varphi}

\def\br{\mathbf{r}}

\def\bp{\mathbf{p}}

\def\bq{\mathbf{q}}

\begin{document}

\title{Quantum  magnetoresistance of Weyl semimetals with strong Coulomb disorder}	
	\author{Ya.~I.~Rodionov}
		\affiliation{Institute for Theoretical and  Applied Electrodynamics, Russian Academy of Sciences, Moscow, 125412 Russia}
	\author{K.~I.~Kugel}
		\affiliation{Institute for Theoretical and  Applied Electrodynamics, Russian Academy of Sciences, Moscow, 125412 Russia}
\affiliation{National Research University Higher School of Economics, Moscow 101000, Russia}
    \author{B.~A.~Aronzon}
	\affiliation{P.~N.~Lebedev Physical Institute, Russian Academy of Sciences, Moscow, 119991 Russia}
	
\date{\today}

\begin{abstract}
We study the effects a strong Coulomb disorder on the transverse magnetoresistance in Weyl semimetals at low temperatures. Using the diagrammatic technique and the Keldysh model to sum up the leading terms in the diagrammatic expansion, we find that the linear magnetoresistance exhibits a strong renormalization due to the long-range nature of the Coulomb interaction
$\rho_{xx} \propto H\ln(eH\hbar v^2/cT^2_{\rm imp}),\ \ \Omega\alpha^{-1/6}\ll T_{\rm imp}\ll \Omega/\alpha^{-3/4}$, where $\Omega = v\sqrt{2eH\hbar/c}$ is the  distance between the zeroth and the first Landau levels, $T_{\rm imp}=\hbar vn^{1/3}_{\rm imp}$ measures the strength of the impurity potential in terms of the impurity concentration $n$ and the Fermi velocity $v$, and $\alpha = e^2/\hbar v$ is the effective fine structure constant of the material. As disorder becomes even stronger (but still in the parametric range, where the Coulomb interaction can be treated as a long-range one), we find that the magnetoresistivity becomes quadratic in the magnetic field $\rho_{xx}\propto H^2$.

\end{abstract}


\maketitle

\section{Introduction}

The discovery of Weyl~\cite{Burkov2011,Ashvin2011,XuScience2015,Weng2015} and Dirac~\cite{LiuScience2014,Neupane2014,Borisenko2014,Jeon_2014} materials opened the door for the study of their transport properties, which are unique due to the {\it relativistic} quasiparticle spectrum characteristic of these materials.

Comprehensive studies show that charged impurities dramatically affect the transport characteristics of Weyl semimetals (WSM) even in the absence of magnetic field~\cite{Skinner2014,Balents2011,Ominato2015,Sarma2015,
Lundgren2014,Ramakrishnan_2015,Rodionov_PRB_15,RodionovPRB2015}.
The magnetic field dependence of numerous Dirac materials is often characterized by large, non-saturating linear magnetoresistance (LMR) that occurs at experimentally-accessible magnetic fields. Most studies agree that disorder is the basis for non-saturating LMR~\cite{KisslingerPRB2017,LeahyPNAS2018} but do not
provide a clear picture of the exact mechanisms that affect it or how to tune it. The recent studies~\cite{NelsonArX2022} demonstrate that the growth of disorder can lead to deviation from the linear behavior of magnetoresistance.  In our paper, we are trying  to elucidate the combined effect of the strong disorder and high magnetic field on the magnetoresistance.

It is worth noting that for the first time the magnetotransport in materials with the Dirac spectrum was studied long before the discovery of Weyl and Dirac semimetals. A.A. Abrikosov in his pioneering paper Ref.~\onlinecite{Abrikosov1998} addressed the transverse magnetoresistance of a gapless semiconductor having linear energy spectrum with charged scatterers in the so-called ultraquantum regime and predicted linear (in the magnetic field) magnetoresistance. The ultraquantum limit implies that the temperature of the compound is much lower than the distance between the zeroth and the first Landau levels (LLs): $T\ll \Omega = v\sqrt{2eH/c}$. The chemical potential $\mu$ is defined by the electroneutrality condition and, in principle, can vary from $\mu<\Omega$ to $\mu\gg\Omega$.

The Abrikosov's treatment is perturbative in the disorder strength, hence  the impurity concentration is assumed to be small enough. The lucky coincidence is that a WSM with Coulomb impurities has an additional small parameter. In the ultraquantum limit, the impurity Debye length $\kappa^{-1}$ (which plays the role of disorder correlation length) is  much larger than the characteristic length of Landau levels (LL) wave function ($l_H/\sqrt{n_0}=\sqrt{c\hbar/eHn_0}$ for the $n_0$th LL): $\kappa l_H/\sqrt{n_0}\ll 1$ for a typical WSM.  Indeed, for all the impurity concentrations under consideration, the Debye screening length is given by the condition
\begin{gather}
 \kappa^{-1}\sim\alpha^{-1/2} l_H/\sqrt{n_0},    \quad \alpha = \frac{e^2}{\hbar v \epsilon},\quad l_H =\sqrt{\frac{c\hbar}{eH}} .
\end{gather}
Here, $\alpha$ is the ``fine structure constant" of the WSM and $\epsilon$ is its dielectric constant. In a reference Dirac semimetal Cd$_3$As$_2$, we have $\alpha \approx 0.05$  ~\cite{LiuScience2014, Aubin1977}. Parameter $n$ is the characteristic LL number contributing to conductivity: $n_0\sim\max\{\Omega^2, W^2,\mu^2\}/\Omega^2$

\begin{figure}[t]
\centering
 \includegraphics[width=0.4\textwidth]{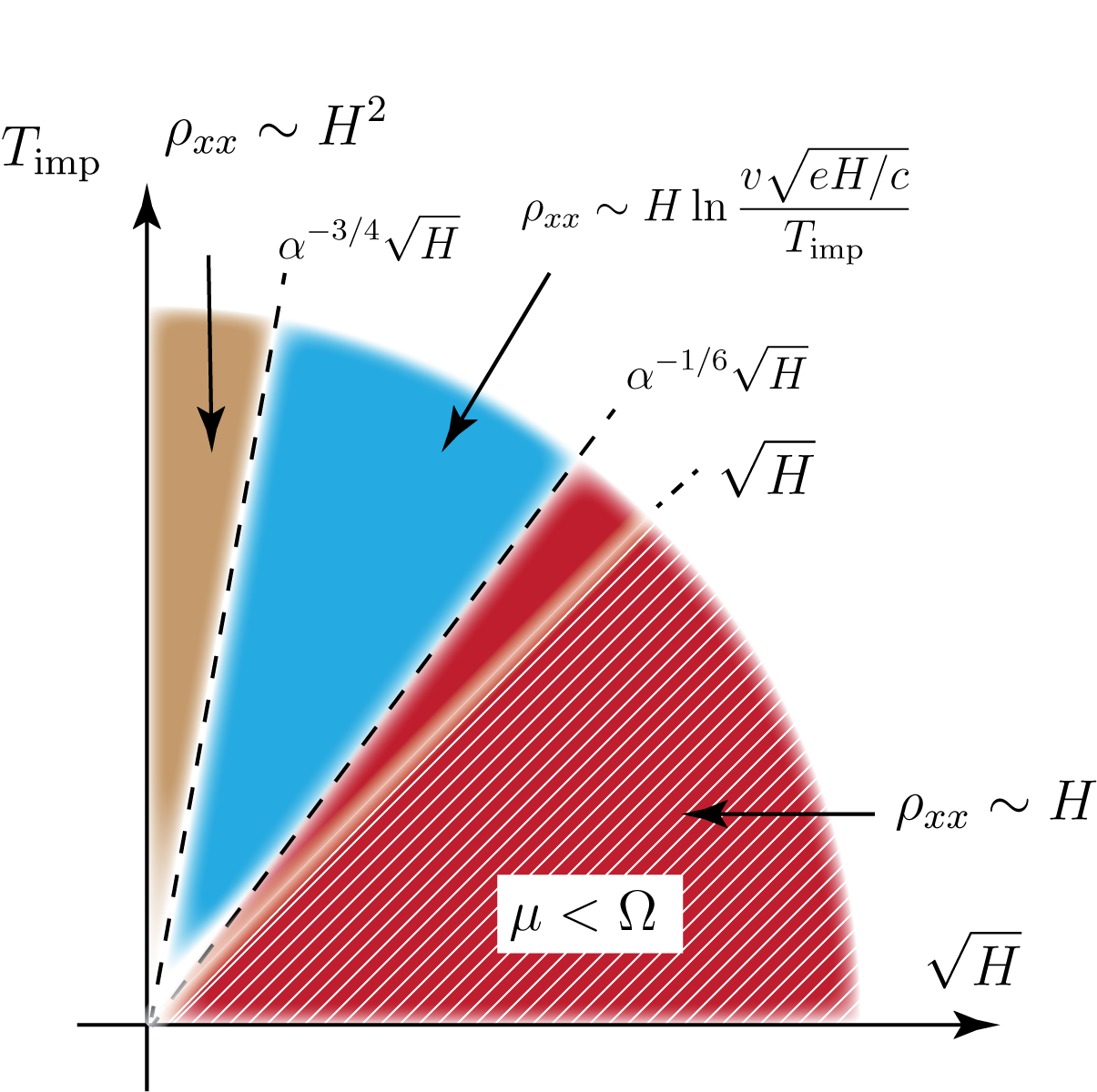}
 \caption{Different parametric regimes for magnetoresistivity $\rho_{xx}$ as a function of the magnetic field $H$ and disorder strength $T_{\rm  imp} = n_{\rm imp}^{1/3}v$. The hatched region represents the regime, where the chemical potential $\mu$ is below the first LL, $\mu\ll\Omega$. The red region represents the previously studied case of a weak disorder. The blue region corresponds to the case of an average disorder. In this regime, the LLs are still well separated. The beige region is the case of a strong disorder. The LLs are smoothed out.}
\label{phase0}
\end{figure}

A small value of the fine structure constant allows us to perform the double perturbative expansion: (i) in the disorder strength, and (ii) in the $\kappa l_H$ parameter.  Recently, the  calculation of transverse magnetoresistance has been extended to higher temperatures and chemical potential, $T, \mu\gg\Omega$, lifting limitations of the ultraquantum limit approach~\cite{KlierPRB2015,KlierPRB2017,LeePRB2017, SongPRB2015}.

However, all the listed papers considered the limit of weak disorder only. The first two papers treated the Coulomb disorder as a modified short-range ($\delta$-functional) interaction. Hence, the corresponding predictions can be considered to be of a qualitative nature.

In this paper, we present the approach that allows us to lift the limitation of weak disorder exploiting its long-range nature. We generalize the previous results for trsansverse magnetoresistance to the case of much stronger disorder, as long as its very strength allows to treat it as a long-range one. Instead of using the perturbation series in disorder amplitude, we formulate the perturbation theory in terms of the $\kappa l_H$ parameter only.

The results of the paper can be summarized in the  phase diagram presented in Fig.~\ref{phase0}. We predict two new parametric regimes, where the magnetoresistance deviates from its linear in the magnetic field dependence.

The linear magnetoresistance sector predicted by Abrikosov~\cite{Abrikosov1998} is painted by the red color in the phase diagram and corresponds to the lowest disorder regime. As disorder strength (measured in Fig.~\ref{phase0} in $T_{\rm imp} = \hbar n_{\rm imp}^{1/3}v$ units) increases, the linear magnetoresistance becomes renormalized by the field dependent $\log$ factor (blue region) Then, it changes its behavior to the $H^2$ dependence. The hatched region in the phase diagram represents the simplest case of a small chemical potential, for which the magnetotransport is entirely determined by charge carriers at the zeroth and first LLs.

The paper is organized as follows. In Section II, we introduce the model and all notation. Section III deals with the averaging technique. In Section IV, we perform the computation of magnetoconductivity. In Section V, we discuss the obtained results.

\section{model} \label{Model}

\subsection{Hamiltonian} \label{Hamiltonian}

We start with the standard Hamiltonian for electrons in Coulomb disorder potential
\begin{gather}
    \label{ham1}
    \begin{split}
       H&=H_0+H_{\rm imp},\\
       H_0&= v\int\psi^\dag(\br)\bm{\sigma}\left(\bp-\frac{e}{c}\bA\right)\psi(\br) d\br,\\
       H_{\rm imp} &= \int \psi^\dag(\br) u(\br)\psi(\br) d\br,
    \end{split}
\end{gather}
where $H_0$ is the Hamiltonian of free non-interacting Weyl fermions, $\psi(\mathbf{r})$ and $\psi^\dag(\mathbf{r})$ are the fermion annihilation and creation operators, $\bm{\sigma}\bp$ is the quasiparticle kinetic energy, and $\bm{\sigma}$ is the pseudospin operator. $H_{\rm imp}$ is responsible for the interaction between electrons and Coulomb impurities. Throughout the paper, we set $\hbar = 1$.

In what follows, we completely discard the quasiparticle scattering between different Weyl nodes in the Brillouin zone. This can be done due to the smoothness and long-range nature of the Coulomb potential created by charged impurities. The tunneling between the Weyl nodes can also be in principle assisted by strong magnetic fields~\cite{Lee2017,Rodionov2018}. Nevertheless, we discard this effect here as well, since it occurs only at extremely high fields.

The disorder potential is assumed to be screened by electrons. Hence, it is given by the profile
\begin{gather}
    u(\mathbf{k}) = \frac{4\pi e^2}{\epsilon}\frac{1}{k^2+\kappa^2} ,
\end{gather}
where $\epsilon$ is the dielectric constant and $\kappa$ is the inverse Debye screening length. It is given by the relation
\begin{gather}
   \label{DebyeLength}
    \kappa^2 = \frac{4\pi e^2}{\ve_\infty}\frac{dn}{d\mu} ,
    \end{gather}
where $n$ and $\mu$ are the electron density and chemical potential, respectively. Here, it is necessary to make the following comment. The thermodynamic density of states depends on the disorder itself and as a result, on $\kappa$. Therefore, Eq.~\eqref{DebyeLength} should be understood as a self-consistency equation on $\kappa$.
If the impurity concentration is not too large in the leading expansion in terms of $l_H\kappa\ll1$, one obtains the following expression for  $\kappa$ (see Appendix \ref{app1}) in the limits of the weak and strong disorder

\begin{gather}
\label{DebyeLength3}
    \kappa^2 = \frac{\alpha}{2\pi v^2}
       \big(\max\{W^2,\Omega^2\}+\mu^2\big),\\
    W^2 = \frac{2\pi n_{\rm imp}\alpha^2 v^2}{\kappa}\label{disorder}.
\end{gather}
Here, $\mu$ is the level of doping of the WSM.
Based on Eq.~\eqref{DebyeLength3}, we discuss the validity of long-range disorder approximation. Suppose that $\Omega\ll W$.  Yet, the applicability of the long-range disorder approximation implies that the disorder correlation length is much larger than the characteristic length of the wave function. In the $W\gg\Omega$ limit, the latter is of the order of $\max\{W,\ \mu\}$. Since the inverse Debye length itself is of the order of $\kappa\sim\sqrt{\alpha} \max\{W, \mu\}$, we come to the conclusion that for the computation of the conductivity, the condition $\alpha\ll 1$ guarantees that the potential can always be considered as a long-range one.

\subsection{Green's functions and conductivity}
\label{Greenfun+cond}

The Kubo formula for the longitudinal conductivity reads
\begin{gather}
\label{conductivity0}
\begin{split}
   \sigma_{xx}= &
      \int\frac{\de f_{\rm F}(\ve)}{\de\ve} d\ve\langle \sigma_{xx}(\ve)\rangle , \\
   \sigma_{xx}(\ve) = &e^2v^2\int \frac{d\bp \; dx^\prime}{(2\pi)^3}\\
    &\times{\rm Tr}\langle\Big\{\sigma_{x}{\rm Im}G^R(x,x^\prime;\ve,\bp)\sigma_{x}{\rm Im}G^R(x^\prime,x;\ve,\bp)\rangle\Big\}.
    \end{split}
\end{gather}
In Eq.~\eqref{conductivity0}, angular brackets mean the averaging over the disorder potential. The integration over momentum  $\mathbf{p}$ is performed in the $yz$ plane. The Green's functions are defined as follows
\begin{gather}
\label{green-zero}
    \begin{split}
    &G^R(x,x^\prime;\ve,\bp)= \sum\limits_{n=0}^\infty S_n(x_{p_y})G_n(\ve,\bp_n) S^\dag_n(x_{p_y}^\prime) ,\\
              &S_n(s)=\begin{pmatrix}
            \chi_n\big(s\big)\ &\ 0\\
            0\ &\ \chi_{n-1}\big(s\big)
           \end{pmatrix},\\
           &G_n(\ve,p_z)=\frac{\ve+v\bm{\sigma}\cdot \mathbf{p}_n}{(\ve+i0)^2-\ve_n^2}, \\
           &x_{p_y}=x-p_y l_H^2.\ \
    \end{split}
\end{gather}
Here, $\chi_n\big(s\big)$ is the normalized oscillator wave function of the $n$th state and
\begin{gather}
\mathbf{p}_n=(0,\sqrt{2n}/l_H,p_z)
\end{gather}
is the effective 2D momentum. In what follows, $G_n(\ve, \mathbf{p}_n)$ will be referred to as \textit{the $n$th component of the Green's function} .

Expressing the trace in formula~\eqref{conductivity0} in terms of the components of the Green's functions, we can rewrite $\sigma_{xx}$ as
\begin{gather}
    \label{cond2}
    \begin{split}
    &\sigma_{xx}(\ve) = e^2\frac{\Omega^2}{2\pi^2}\\
    &\times\sum\limits_{n}
    \int\frac{dp_z}{2\pi}\Big[\langle{\rm Im}G^R_{n,11}(\ve,p_z)
    {\rm Im}G^R_{n+1,22}(\ve,p_z)\\
    &+{\rm Im}G^R_{n,12}(\ve,p_z){\rm Im}G^R_{n+1,12}(\ve,p_z)\rangle\Big].
    \end{split}
\end{gather}
The last line in Eq.~\eqref{cond2} appears owning to  the disorder vertex corrections only and, as is proven in Appendix B, vanishes in the ultraquantum limit.

\subsection{Chemical potential}

Before proceeding any further, it is important to comment on a possible relation between impurity concentration and the density of electrons.  The density $n(\mu)$ of excess electrons depends on the density of donor $n_D$ an acceptor $n_A$ impurities as follows
\begin{gather}
     n_D-n_A = n(\mu)
\end{gather}
due to electroneutrality condition.
The impurity concentration is, however, given by the corresponding sum $n_{\rm imp} = n_D + n_A$. Therefore, it is important to distinguish between two situations: (i) compensated WSMs, where $n_D-n_A\ll n_D+n_A$ and the electron concentration $n$ is independent of the full impurity concentration and (ii) uncompensated ones, when $n \sim \max\{n_D, n_A\}\sim n_{\rm imp}$. In this paper, we plan to compare our results with the experiment~\cite{NelsonArX2022}, where $n_A\sim n_{D}\sim n_{\rm imp}$, and the chemical potential becomes a function of the full impurity concentration.

In what follows, we always take into account the the finite chemical $\mu$ potential of the electrons of WSM in the computation of $\sigma_{xx}$.
The case of compensated WSM is considered for completeness for the case of small impurity concentration.

\section{Averaging over the disorder potential}
\label{disorder-aver}

\subsection{Keldysh model}

Our goal is to study the effect of a strong, but long-range disorder.
First of all, let us write the disorder correlation function
\begin{gather}
\begin{split}
    g(\bq) \equiv n_{\rm imp}|u(\bq)|^2 =  \frac{8\pi\kappa W^2}{(\bq^2+\kappa^2)^2},
\end{split}
\label{disorder_cor}
\end{gather}
where $W^2$ is defined in Eq.~\eqref{disorder} and $u(\mathbf{q})$ is the screened Coulomb potential.

To study the effects of strong, but smooth disorder, the Keldysh model seems to be the most appropriate tool. A good description of this model can be found in Refs.~\onlinecite{SadovskiiBook_Diagram2019,EfremovArxiv2022}.

Let us outline the main blocks of this model, on which we base our further discussion. The principal idea is that due to long-range nature of the disorder, its  correlation function is strongly peaked at $q=0$ in the momentum space having the characteristic scale of the wave function (in the momentum representation, this scale is $\max\{\mu, \Omega, W\}/v$ ). The main simplification of the Keldysh model comes from the substitution of the disorder correlation function with $\delta$-function in the momentum space. Thus, it completely discards the momentum transfer related  the disorder in all terms of perturbation series
\begin{gather}
    g(\bq)\Rightarrow W^2\delta(\bq).
\end{gather}
Therefore, we take the limit $\kappa l_H\rightarrow 0$, while keeping parameter $W^2$ finite.

Here, we have to be careful, since the small parameter $\kappa l_H$ is proportional to the impurity concentration itself. So, the legitimate question is:
\textit{can we consider a strong disorder and still treat it as a long-range one?} To understand this, let us answer the question, what do we mean speaking about a weak and strong disorder in our problem.

\subsubsection{Disorder regimes}

The previous Abrikosov's treatment of the disorder took it as the smallest energy scale of the problem $W\ll\kappa v$. From that, we immediately conclude
\begin{gather}
    T_{\rm imp}\ll\frac{\Omega}{\alpha^{1/6}},\quad\hbox{Abrikosov's limit}.
\end{gather}
In our treatment, we roughly split the disorder strength into two regimes according to the structure of the perturbation series.

The \textit{average} disorder strength is defined by the condition $\kappa v\ll W\ll\Omega$. The \textit{strong} disorder corresponds to the case $W\gg \Omega$. If the WSM is compensated, the doping level and impurity concentration are unrelated, and it is possible to have a situation of a small chemical potential $\mu <\Omega$ and any disorder strength.

In terms of $T_{\rm imp}$, we have the following condition for compensated WSM (see estimates in the Appendix A).
\begin{center}
\begin{tabular}{|c|| c |c|}
WSM  & average dis. & strong dis.\\
\hline
 \ & \ &\ \\
compensated & $\frac{\Omega}{\sqrt{\alpha}}\gg T_{\rm imp}\gg\frac{\Omega}{\alpha^{1/6}}$ & $T_{\rm imp}\gg\frac{\Omega}{\sqrt{\alpha}}$\\
 \ & \ &\ \\
uncompensated & $\frac{\Omega}{\alpha^{3/4}}\gg T_{\rm imp}\gg\frac{\Omega}{\alpha^{1/6}}$ & $ T_{\rm imp}\gg\frac{\Omega}{\alpha^{3/4}}$\\
 \ & \ &\ \\
\end{tabular}
\end{center}

\subsubsection{Keldysh conductivity}
The Keldysh model allows us to sum up exactly all terms of the perturbation series to find out the exact expression for the system Green's function
\begin{gather}
    \label{green_exect}
    G^R = \frac{1}{W\sqrt{2\pi}} \int\limits_{-\infty}^\infty e^{-\mu^2/2W^2}\frac{d\mu}{G^{-1}_{R0}-\mu}.
\end{gather}
Here, $G_{R0}$ is the unperturbed Green's function of the theory.
Formally, the exact averaging over the disorder potential in \eqref{green_exect} is equivalent to the introduction of the Gaussian fluctuating chemical potential with the dispersion equal to the disorder strength $W^2$.
The reader is encouraged to convince him(her)self that the full summation of disorder lines for conductivity expression~\eqref{conductivity0} including the summation of all vertex diagrams leads to the following expression
\begin{gather}
\label{condKeldysh}
\begin{split}
   &\sigma_{xx}=
     \int d\ve\frac{d f_{\rm F}(\ve-\mu)}{d\ve}\frac{dE}{W\sqrt{2\pi}}e^{-E^2/2W^2}
     \sigma_{xx}(\ve-E).
    \end{split}
\end{gather}
Changing the Fermi function derivative to the $\delta$-function in the limit $T\rightarrow 0$, we immediately obtain a simplified expression
\begin{gather}
\label{condKeldysh1}
\begin{split}
   &\sigma_{xx}=
     \int \frac{dE}{W\sqrt{2\pi}}e^{-(E-\mu)^2/2W^2}
     \sigma_{xx}(E).
    \end{split}
\end{gather}
In the latter expression, the conductivity appearing in the r.h.s. should be taken with non-interacting Green's functions. (The disorder averaging is already taken into account by the Gaussian contraction).

However, the immediate plugin of the Green's function \eqref{green_exect} into~\eqref{condKeldysh1} yields zero result. This happens due to the fact, that, as seen from~\eqref{cond2}, the Kubo formula always mixes Green's function components from $n$th and $n+1$th LLs. Let us elaborate on some details.

\subsection{Corrections to the Keldysh model}

The disorder averaging can be split into two parts: the independent averaging of separate Green's functions entering the Kubo loop and the so called vertex diagrams (when disorder line connects two Green's functions belonging to opposite sites of the loop). If the momentum transmitted by disorder is zero, both types of averaging are treated on equal footing leading to formula~\eqref{condKeldysh}.

However, returning to large but finite correlation length $\kappa l_H\neq 0$ reveals a more subtle picture.

As the analysis shows (see Appendix B), the diagrams containing vertex corrections vanish. As a result, the first correction to the Keldysh model is as follows. Both Green's functions entering the Kubo expression need to be averaged independently.

Therefore, the corrected expression for the conductivity has the form
\begin{gather}
    \label{condKeldysh2}
    \begin{split}
    \sigma_{xx}(\ve)& = e^2\frac{\Omega^2}{2\pi^2}\\
    &\times
    \sum\limits_n\int\frac{dp_z}{2\pi}{\rm Im}\langle G^R_{n,11}(\ve,p_z)\rangle
    \langle{\rm Im}G^R_{n+1,22}(\ve,p_z)\rangle ,
    \end{split}
\end{gather}
where the imaginary part of the averaged Green's function is given by the following expression
\begin{gather}
\label{GreenKeldysh}
    \begin{split}
   {\rm Im} G^R_{n,\alpha\alpha}(\ve,p_z)&=-\sqrt{\frac{\pi}{2}}\frac{1}{W}
    \bigg\{\Big(\frac{1}{2}-\frac{p_z}{2\ve_{n}}\Big) e^{-(\ve - \ve_n(p))^2/2W^2}\\ &+\Big(\frac{1}{2}+\frac{p_z}{2\ve_n}\Big)e^{-(\ve + \ve_n(p))^2/2W^2}\bigg\}.
    \end{split}
\end{gather}
The further analysis shows the necessity to distinguish the computation between the cases of average and strong disorder. The situation in these limits is different due to the following reasons. In the small disorder limit $W\ll\Omega$, as we see from expression~\eqref{GreenKeldysh}, the spectral weight (${\rm Im}\, G$) is suppressed as a function of energy away from the Green's function {\it mass shell} $\ve = \ve_n$ (the width of the function is $W$).

\begin{figure}[h]
\centering
 \includegraphics[width=0.4\textwidth]{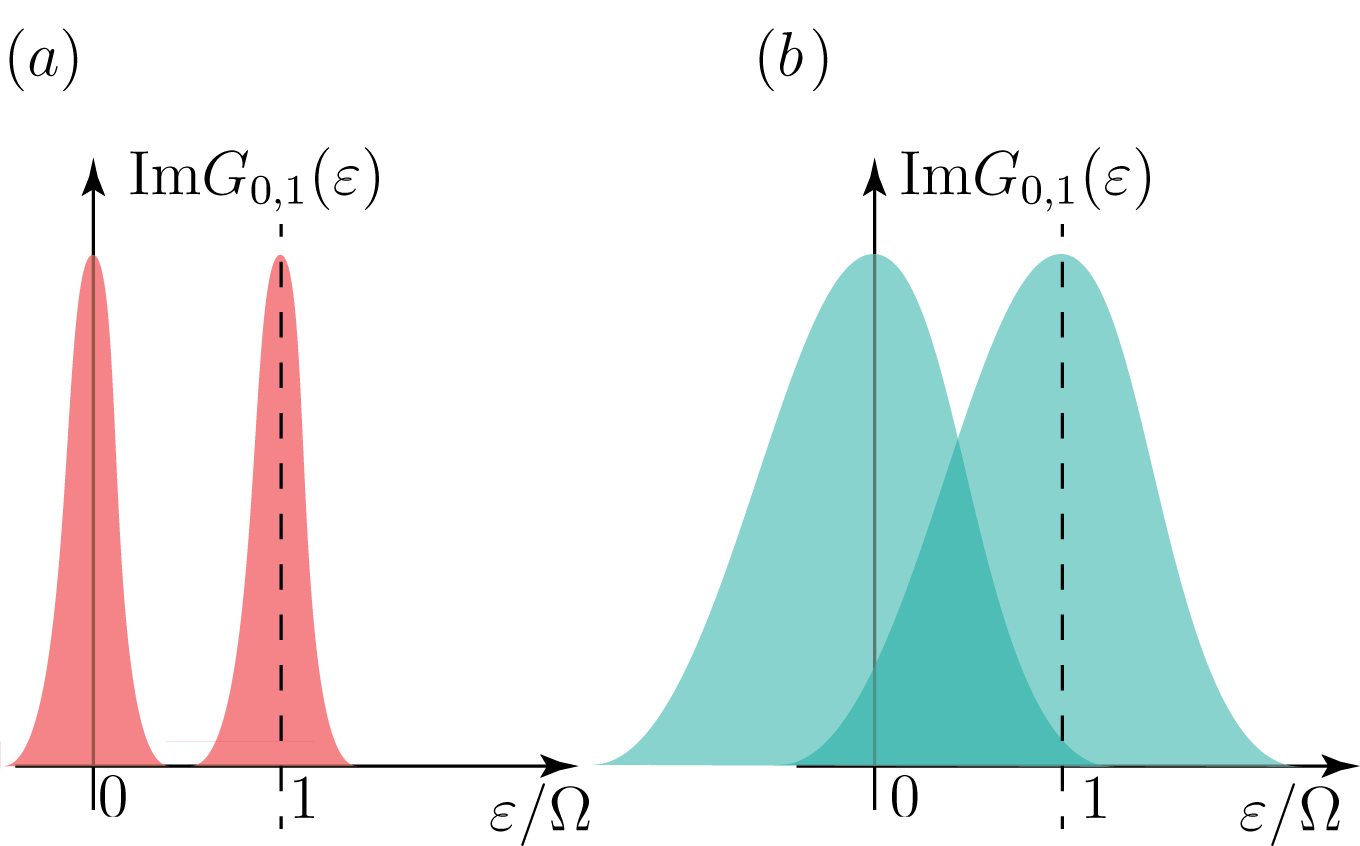}
 \caption{
 The spectral weight ${\rm Im}G_{n}$ of the Green's function components for the zeroth and the first LLs.
 (a) Weak disorder, $W\ll\Omega$;
 (b) Strong disorder, $W\gg\Omega$.}
\label{keldysh0}
\end{figure}

The situation for different disorder strengths is illustrated  in  Fig.~\ref{keldysh0}. The case of average disorder corresponds to exponentially small overlap between the Green's functions entering the Kubo formula. The computation of conductivity in this case requires yet another elaboration of the Keldysh model.

However, the strong disorder case is treated in the easiest way, since the overlap is significant and one can proceed along the lines of Eq.~\eqref{condKeldysh} with the Green's functions given by~\eqref{GreenKeldysh}.

\subsubsection{Conductivity, strong disorder $W\gg\Omega$}

As shown in Appendix A, the electron doping level is related to the disorder strength as follows: $\mu\sim W/\alpha^{3/4}\gg W$ (see Eq.~\eqref{doping_strong1}). We can use this to our advantage for deriving an approximate formula for the conductivity. Performing integration over momentum $p_z$  in~\eqref{condKeldysh2} with the Green's functions from \eqref{GreenKeldysh}, we obtain the following expression for conductivity (see Appendix B for details)
\begin{gather}
\label{cond_large}
    \sigma_{xx}=\frac{e^2\mu^2}{12\pi^{3/2} W},\ \ W\gg\Omega.
\end{gather}
We see that the conductivity $\sigma_{xx}$ saturates at high disorder levels. As we will see in Section~\ref{section_magneto}, this leads to quadratic field dependence of magnetoresistance.

\subsubsection{Conductivity, average disorder $\kappa v\ll W\ll\Omega$}

We see that if $W\ll \Omega$, the Green's functions in~\eqref{condKeldysh2} have exponentially small overlap (see Fig.~\ref{keldysh0}b). This means that even a separate averaging of the Green's function according to our method of improving the Keldysh model is still a too rough approximation as it leads to the exponentially small overlap of Green's functions spectral weight and is exponentially suppressed at $(\sim e^{-\Omega/W})$. The problem is that the approximation of the infinite disorder correlation length $(\kappa l_H\rightarrow 0)$ completely neglects the possible mixing of $n$ and $n+1$ components of the Green's function due to the very framework of the Keldysh model. The main contribution to the conductivity comes, as we are going to see, from the  $\kappa l_H$ expansion, and it is non-exponential.

\begin{figure}[h]
\centering
 \includegraphics[width=0.4\textwidth]{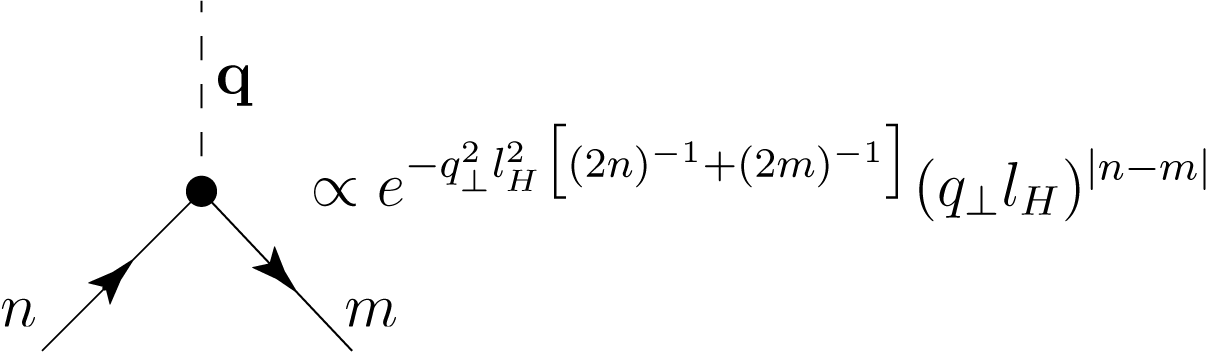}
 \caption{
 The disorder vertex. Here $\mathbf{q}_\bot = (q_x, q_y).$}
\label{dis0}
\end{figure}

If we allow for the finite momentum transfer due to disorder interaction, the structure of the disorder interaction, as seen from the definition of the Green's function~\eqref{green-zero}, reads (see Fig.~\ref{dis0})
\begin{gather}
    \begin{split}
    V_{nm}(\mathbf{q}) = \int\limits_{-\infty}^\infty dx e^{iq_x x}\chi_n(x)\chi_m(x-q_y  l_H^2)
    \\
    \propto e^{-q_\bot^2 l_H^2/4}|q_\bot l_H|^{|n-m|}.
    \end{split}
\end{gather}
Therefore, at finite transmitted momenta $q$, the disorder mixes different components of the Green's functions.

\subsection{Perturbation series at a finite correlation length of disorder}
\begin{figure}[h]
\centering
 \includegraphics[width=0.4\textwidth]{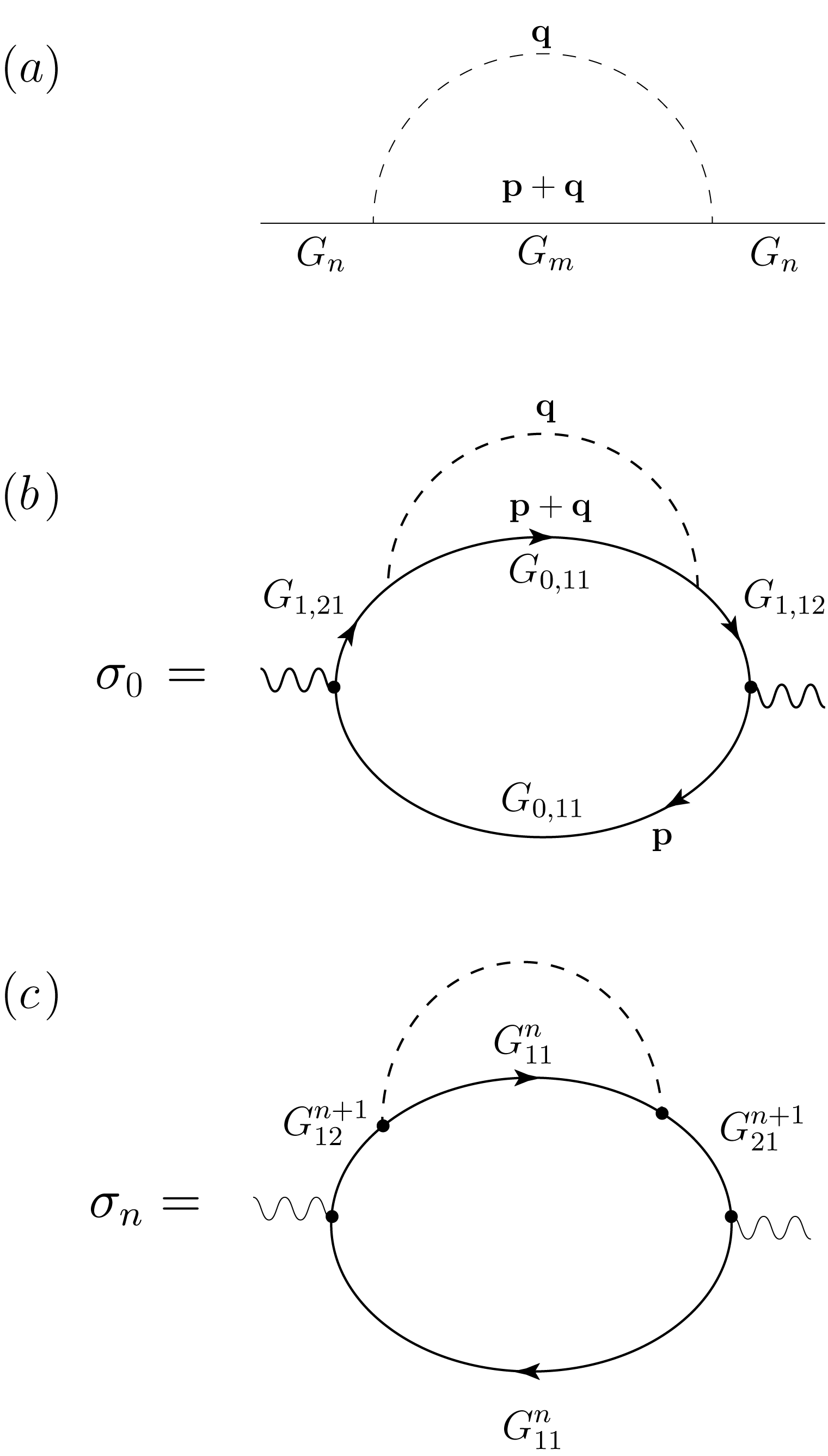}
 \caption{(a) Correction to the Green's function in the first order in the disorder strength;  (b) The lowest in disorder nonvanishing term  contributing to the conductivity;  (c) Contribution to the conductivity from higher LLs ($n>1$).}
\label{abrikosov0}
\end{figure}

From~\eqref{cond2}, we see that the conductivity always contains parts of Green's functions corresponding to the adjacent LLs.
The corresponding correction to the Green's function is illustrated in Fig.~\ref{abrikosov0}a.

The disorder broadens the imaginary part of the Green's function and  mixes the $n$th component of the Green's function with other components.

Here, we need to distinguish between two different disorder strengths:

(i) weak disorder, $W\ll v\kappa $ (the Abrikosov's limit, red region   $T_{\rm imp} \ll \alpha^{-1/6}\Omega$ in the phase diagram~Fig.~\ref{phase0})

(ii) average disorder strength, $ v\kappa \ll W \ll \Omega$ (blue region
$ \alpha^{-1/6}\Omega \ll T_{\rm imp} \ll \alpha^{-3/4}\Omega$).

The Abrikosov's limit, in  turn, is split into other two cases.
If the doping level is positioned below the first LL, $\mu<\Omega$,
only the zeroth and the first LLs contribute to the conductivity. This is possible for an uncompensated WSM, only if $T_{\rm imp}<\Omega$ (the shaded region of the phase diagram~\ref{phase0}). For higher doping levels, we are going to observe Shubnikov--de Haas oscillations due to contributions from higher LLs.

In Appendix B, we argue that for $\mu<\Omega$, the leading term in \eqref{cond2} comes from $n=0$ component of the Green's function and the corresponding diagram is presented in Fig.~\ref{abrikosov0}b. The expression for conductivity in the limit $W\ll\Omega$ thus reads
\begin{gather}
    \label{cond3}
    \begin{split}
    \sigma_{xx}(\ve) = e^2\frac{\Omega^2}{2\pi^2}
    \int\frac{dp_z}{2\pi}{\rm Im}G^R_{0,11}(\ve,p_z)
    {\rm Im}G^R_{1,22}(\ve,p_z),
    \end{split}
\end{gather}
which leads to the well-known Abrikosov's answer for the conductivity:
\begin{gather}
\label{cond_abr}
    \sigma_{xx} = \frac{e^2}{8\pi^3}\frac{v\kappa W^2}{\Omega^2}\ln\frac{1}{\kappa l_H}.
\end{gather}

\section{Average disorder, $v\kappa\ll W\ll\Omega$}

The stronger disorder corresponds to the parametric regime $\kappa v\ll W\ll\Omega$, where the Abrikosov's approach is no longer valid. Indeed, the latter assumes that the broadening of the Green's function $\delta \ve \sim \kappa v$, while a simple estimate from the Keldysh model tells that the spectral width of the Green's function is $\delta\ve\sim W\gg\kappa v$ (see. Eq.~\eqref{GreenKeldysh}).  Therefore, the Keldysh model seems to be an appropriate approach to employ. Indeed, the disorder correlation length $\kappa^{-1}\gg l_H$ is the largest length scale in the system. Let us first keep the correlation length finite and investigate the structure of the self-energy in the perturbative regime.

It is straightforward to check that the diagrams with crossings lead to subleading contributions to the self energy comparing to the diagrams without crossings. The relation has the following form (see Appendix B):
\begin{gather}
    \Sigma^1_n \simeq (\kappa l_H)\Sigma_n.
\end{gather}

\begin{figure}[h]
\centering
 \includegraphics[width=0.4\textwidth]{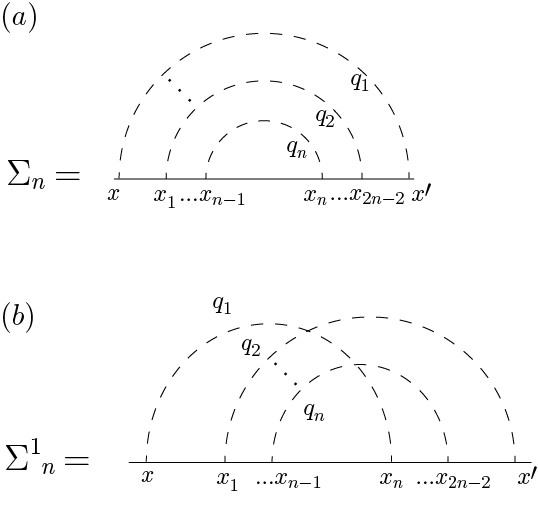}
 \caption{(a) Leading contribution to the self energy for low Landau levels; (b) Subleading contribution.}
\label{abrikosov2}
\end{figure}
Now we proceed to compute the conductivity $\sigma_{xx}$ in the limit of small and large doping $\mu$.

\subsection{Small doping $\mu<\Omega$}
In this case, as was argued before, only the zeroth and the first LLs give contribution to conductivity.
However small doping also implies that the WSM is highly compensated, $|n_D-n_A|\ll |n_D+n_A|$, and the electron density and total impurity concentration are decoupled (indeed, for an uncompensated WSM, the doping level obeys the relation $\mu\sim T_{\rm imp}\gg \Omega$, see Eq.~\eqref{app_density__weak_high}).
This means the phase diagram~\ref{phase0} doesn't describe this case. However, this case is still very important from the experimental viewpoint.

\begin{figure}[h]
\centering
 \includegraphics[width=0.3\textwidth]{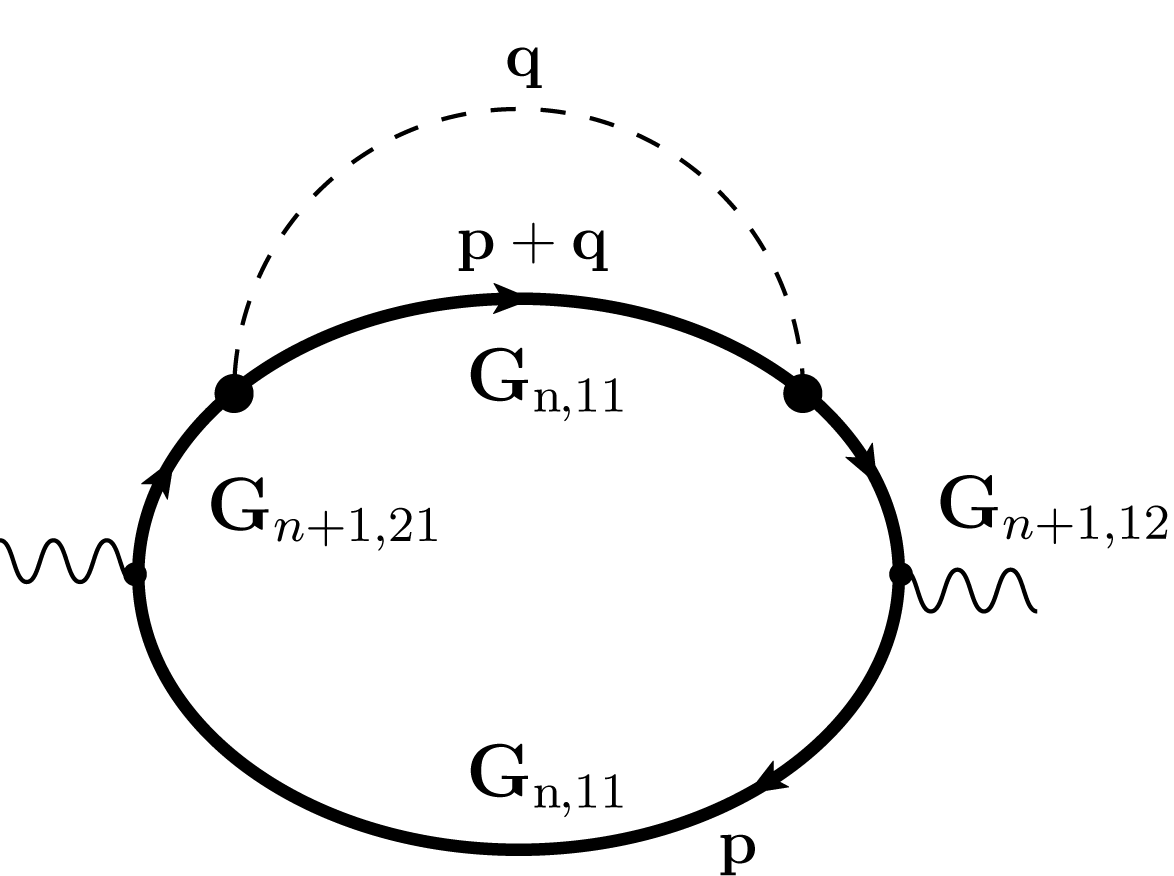}
 \caption{Kubo loop determining the conductivity for arbitrary doping $\mu$.}
\label{skeletal2}
\end{figure}

Therefore, we proceed in the following way.
In the lowest order in expansion parameter $\kappa l_H$, the conductivity is given by the skeletal diagram in Fig.~\ref{skeletal2} with $n=0$.

All Green's functions entering the diagram shown in Fig.~\ref{skeletal2} should be calculated exactly in the $\kappa l_H\ll 1$ limit, i.e. using the Keldysh model. The computation of the self-energy is presented in Appendix B4. The final expression is given by the integral with the following result
\begin{gather}
\begin{split}
\label{cond_fin}
     \sigma_{xx}(\mu<\Omega)&\approx \frac{e^2}{2\pi^2}
     \sqrt{\frac{2}{\pi}}
     \frac{W^2\kappa}{16\Omega^2}
     \int\limits_{-\infty}^\infty\frac{ds}{2\pi}e^{-s^2/2}\ln\frac{\Omega^2}{W^2 \max\{s^2,1\}}\\
     &= \frac{e^2W^2\kappa}{16\pi^2\Omega^2}\ln\frac{\Omega^2}{W^2}.
\end{split}
\end{gather}
Taking into account that $W^2\kappa\sim T_{\rm imp}^3$ (Eq.~\eqref{disorder}) is magnetic field independent, we conclude
that the conductivity in the case of stronger disorder acquires an additional slow logarithmic magnetic field dependence apart from $1/\Omega^2$ dependence in  the Abrikosov's result~\eqref{cond_abr}.

\subsection{High doping, $\mu\gg\Omega$}

If the WSM is not compensated, we have $n_{\rm imp}\sim n$.
This means that the chemical potential is no longer restricted to the region below the first LL. We remember that $\mu\sim T_{\rm imp}$ while the Debye length is $\kappa v \sim\sqrt{\alpha}\mu \sim \sqrt{\alpha}T_{\rm imp}$  (see Eq.~\eqref{kappa_small}). The disorder strength $W\sim\alpha^{3/4}T_{\rm imp}\ll\kappa v$, therefore, in the uncompensated case, the $\mu\gg\Omega$ condition automatically entails $W\ll\kappa v$. The last condition as it may seem, returns the calculation to the Abrikosov's limit. However, we are going to see that the field dependent log renormalization of the conductivity as in the case of compensated WSM described by Eq.~\eqref{cond_fin} persists.

\begin{figure}[h]
\centering
 \includegraphics[width=0.4\textwidth]{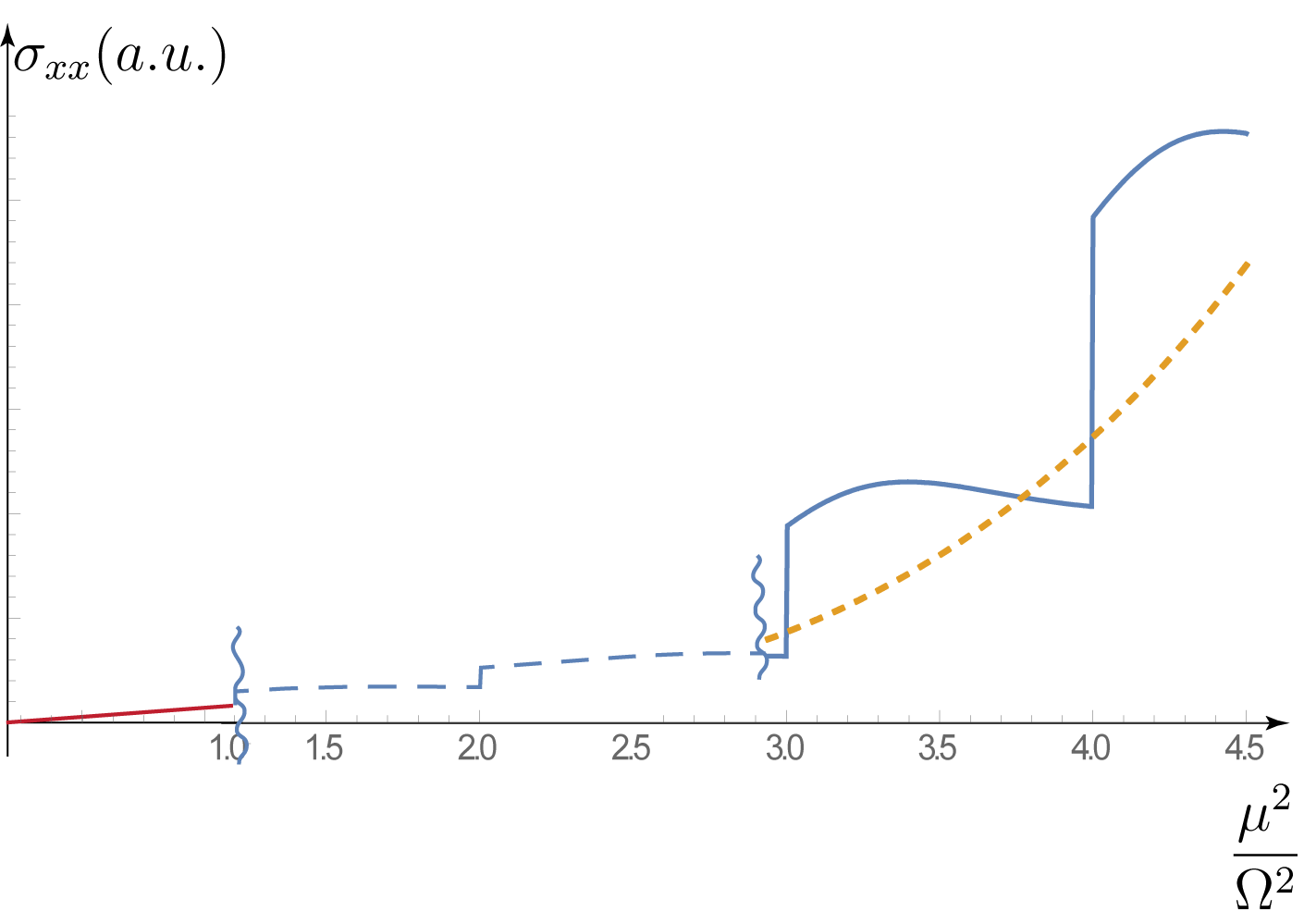}
 \caption{
 The conductivity of WSM $\sigma_{xx}$ for $\mu\gg \Omega$ and $\mu<\Omega$. The enveloping curve (the orange dashed line) is $\sim (\mu^2/\Omega^2)^4\ln\frac{\Omega^2}{W^2}$. The linear part $(\mu<\Omega)$ is depicted in red.
 The blue dashed part corresponds to the conductivity in the intermediate regime $\mu\gtrsim\Omega$ interpolating between two computed cases.
 }
\label{cond2a}
\end{figure}

The conductivity is calculated similar to the case of small doping. The only difference is that the leading contribution comes from the self-energy mixing the $n$th and $(n+1)$th LLs (see Fig.~\ref{skeletal2}). It turns out that for higher LLs, the diagrams without crossings provide the leading contribution as in the case of small doping. This leads to nearly identical calculation.
The result is obtained in Appendix and can be written as
\begin{gather}
    \label{cond_high_mom}
    \sigma_{xx} = \sigma_{xx}(\mu<\Omega)
    \sum\limits_{n=1}^{\frac{\mu^2}{\Omega^2}-1}
    \frac{(n+1)^2\mu}{\sqrt{\mu^2-n\Omega^2}},\ \ \mu\gg\Omega,\ \frac{\mu^2}{\Omega^2}\notin \mathbb{N}.
\end{gather}
The corresponding dependence is presented in Fig.~\ref{cond2a}.

The result~\eqref{cond_high_mom} is the direct extension of the Eq.~\eqref{cond_fin}  multiplied by the Shubnikov--de Haas steps. We see that even in the case of large doping $\mu\gg\Omega$, the conductivity exhibits the strong log renormalization due to the long-range nature of Coulomb disorder.

Our equations \eqref{cond_large}, ~\eqref{cond_fin} and \eqref{cond_high_mom} are the main findings of the paper. They generalize the previous findings for the conductivity for the cases of a strong $W\gg\Omega$ and average disorder $\kappa v\ll W\ll\Omega$.
Now, we would like to express the results in terms of the experimentally relevant magnetoresistivity.

\section{Magnetorestivity, strong disorder}
\label{section_magneto}
The magnetoresistivity $\rho_{xx}$ is given by the following relation
\begin{gather}
\label{magneto0}
    \rho_{xx}=\frac{\sigma_{xx}}{\sigma_{xx}^2+\sigma_{xy}^2}.
\end{gather}
Therefore, to obtain $\rho_{xx}$, we also need to compute the Hall conductivity.
The magnetoresistivity is most interesting in the case of a strong disorder $W\gg\Omega$. As we are going to see, strong disorder smear out Shubnikov--de Haas oscillations turning magnetoresistivity into the nonlinear monotonous function of the applied magnetic field.

The expression for  $\sigma_{xy}$ has a regular nonzero limit at $\kappa l_H\rightarrow 0$ at any value of the disorder strength. This allows us to compute it using the Keldysh model without modifications
\begin{gather}
\label{conductivity1}
\begin{split}
   &\sigma_{xy}=
     \int d\ve\frac{d f_{\rm F}(\ve-\mu)}{d\ve}\frac{dE}{W\sqrt{2\pi}}e^{-E^2/W^2}
     \sigma_{xy}(\ve-E) .
    \end{split}
\end{gather}
Here, $\sigma_{xy}(E)$ is a conductivity, obtained with non-interacting Green's functions.

In our case, $T\rightarrow 0$, and we perform energy integration in \eqref{conductivity1} by setting $\ve=\mu$. This way, we obtain a simplified expression for the conductivity
\begin{gather}
\label{conductivity2}
\begin{split}
   &\sigma_{xy}=
     \int\frac{dE}{W\sqrt{2\pi}}e^{-(E-\mu)^2/W^2}
     \sigma_{xy}(E) .
    \end{split}
\end{gather}

As is known, the Hall conductivity $\sigma_{xy}$ can be split into normal and anomalous part as follows
\begin{gather}
\label{sigma_xy0}
    \begin{split}
    \sigma_{xy} &= \sigma_{xy}^{I}+\sigma_{xy}^{II},\\
    \sigma_{xy}^{I}(E) &=\frac{e^2\Omega^2}{2\pi^2} \sum\limits_{n}\int\frac{dp_z}{2\pi}\int \Big[{\rm Re} G^{R}_{n+1,22}{\rm Im}G^{R}_{n,11}\\
    &
    -{\rm Re} G^{R}_{n,11}{\rm Im}G^{R}_{n+1,22}\Big]
    , \\
    \sigma_{xy}^{II}& = ce\frac{\de }{\de H}n.
    \end{split}
\end{gather}

We show in Appendix C that in the case of strong disorder $\sigma_{xx}^{I} = \sigma_{xx}^{II}$ and we have

\begin{gather}
\label{Hall1}
    \sigma_{xy}\approx\frac{e^2}{3\pi^2 v}\frac{\mu^3}{\Omega^2}.
\end{gather}
As before, the electroneutrality condition gives us $\mu\sim T_{\rm imp}$ (see~\eqref{doping_strong1} in Appendix A). This immediately yields $W\sim \alpha^{3/4}T_{\rm imp}$.
Comparing ~\eqref{Hall1} and ~\eqref{cond_large}, we see that the Hall conductivity dominates over the longitudinal one
\begin{gather}
    \frac{\sigma_{xy}}{\sigma_{xx}}\sim\frac{T_{\rm imp} W}{\Omega^2}\gg\frac{1}{\alpha^{3/4}}\gg 1.
\end{gather}
Therefore, we discard $\sigma_{xx}$ in the denominator of Eq.~\eqref{magneto0}
to obtain
\begin{gather}
\label{rho_large}
    \rho_{xx} =  \frac{\sigma_{xx}}{\sigma^2_{xy}}=\frac{3\pi^{5/2}}{4\alpha}\frac{\Omega^4}{\mu^4W}\propto\frac{\Omega^4}{T_{\rm imp}^5}\sim H^2.
\end{gather}

Therefore, magnetoresistance~\eqref{rho_large} exhibits the quadratic magnetic field dependence for a strong disorder (beige region of the phase diagram).

\section{Discussion and comparison with other limits}

The theoretical study of the  magnetoconductivity in the ultraquantum limit with large Coulomb disorder has not been addressed yet, therefore it is somewhat challenging to compare our results with other theoretical works, since there are very few.
However,  some qualitative analysis of the magnetoconductivity with relatively strong Coulomb disorder was undertaken in the Ref.~\onlinecite{KlierPRB2015}. We stress however, that in the latter paper the Coulomb disorder was
treated as extremely short-range (point-like), with field dependent amplitude.

The authors of Ref.~\onlinecite{KlierPRB2015} found that the conductivity saturates to the field independent limit at
large disorder strength. They found that $\sigma_{xx}\sim \alpha T_{\rm imp}$. We find a similar result (see Eq.~\eqref{cond_large}). In this case, $W\sim\alpha^{3/4}T_{\rm imp}$, and the magneto conductivity reads
\begin{gather}
    \sigma_{xx} \sim \alpha^{1/4}T_{\rm imp}.
\end{gather}

Also, we need to take into account the upper limit on the disorder strength of our study. The characteristic scale of the wave function $W$ should be smaller than the bandwidth  of the WSM.

Our results can be matched with experimental study in Ref.~\onlinecite{NelsonArX2022} undertaken for ${\rm Cd}_3{\rm As}_2$. In this work, for uncompensated samples with impurity concentration $n_{\rm imp}\sim 10^{17}\ {\rm cm}^{-3}$, the magnetic field $H\sim 1\ \rm T$ and Fermi velocity $v\sim 10^6\ {\rm m/s}$, we obtain $T_{\rm imp}/\Omega\approx 1.16$. Therefore, we expect, that at this limit, the field dependence of magnetoresistance should fall into the blue region of our phase diagram; this means that $\rho_{xx}\propto H\ln H$ rather than $H^2$. We predict that the quadratic regime in the experiment reported in Ref.~\onlinecite{NelsonArX2022}  starts at $H\sim 0.02\ \rm T$.

Conductivity~\eqref{cond_fin} in the limit of average disorder strength $\kappa v\ll W\ll \Omega$ should be matched with the Abrikosov's linear magnetoconductivity at $W\sim \kappa v$. Comparing it with Eq.~\eqref{cond_abr}, we see that this  is indeed the case if we put $\ln(\Omega/W)\sim 1$ (we keep in mind that the computation of~\eqref{cond_fin} is performed with the log-accuracy, and the correspondence, of course, is up to numerical factor).

Concluding, we studied for the first time the transverse magnetoconductivity of the WSM in a non-perturbative regime with respect to disorder strength. Our findings generalize previous results and match the existing ones for small disorder strengths.

\section*{Acknowledgements}
The work was supported by the Russian Science Foundation (Project No. 21-12-00254, https://rscf.ru/en/project/21-12-00254/). The work of Ya.I.R in the part concerning the numerical calculations was supported by the Russian Foundation for Basic Research (project No. 20-02-00015).

\appendix
\section{Electron density}
\label{app1}
The excess particle density can be computed in all orders of the perturbation theory in disorder strength using the Keldysh model
\begin{gather}
\begin{split}
    &n = \mathcal{N}(\mu)-\mathcal{N}(0),\\
    &\mathcal{N}(\mu) =  -\int  n_{\rm F}(\ve,\mu){\rm Im}\langle G^R(x,x,\ve,\bp)\rangle_{\rm dis}\frac{d^2pd\ve}{4\pi^3}.
\end{split}
\end{gather}
Here, $n_{\rm F}(\ve,\mu) = (e^{(\ve-\mu)/T}+1)^{-1}$ is the Fermi function.
Slightly transforming the formula and plugging in the Keldysh Green's functions as well as setting $T=0$, we arrive at
\begin{gather}
\label{number}
    n =\frac{1}{\sqrt{2\pi}W}\frac{\Omega^2}{2\pi^2 v^2} \int\limits_{-\mu}^\mu d\ve\int\frac{dp_z}{2\pi}\sum\limits_{n = 0}^{\infty\prime} e^{-(\ve-\ve_n)^2/2W^2}.
\end{gather}

Here, the prime in the sum means that the term with $n = 0$ should be taken with coefficient $\frac{1}{2}$.
The last expression
is easily analyzed in two important limits: (i) $W\gg\Omega$ and  (ii) $W\ll\Omega$.

\subsection{Strong disorder $W\gg\Omega$}
\begin{figure}[H]
\centering
 \includegraphics[width=0.15\textwidth]{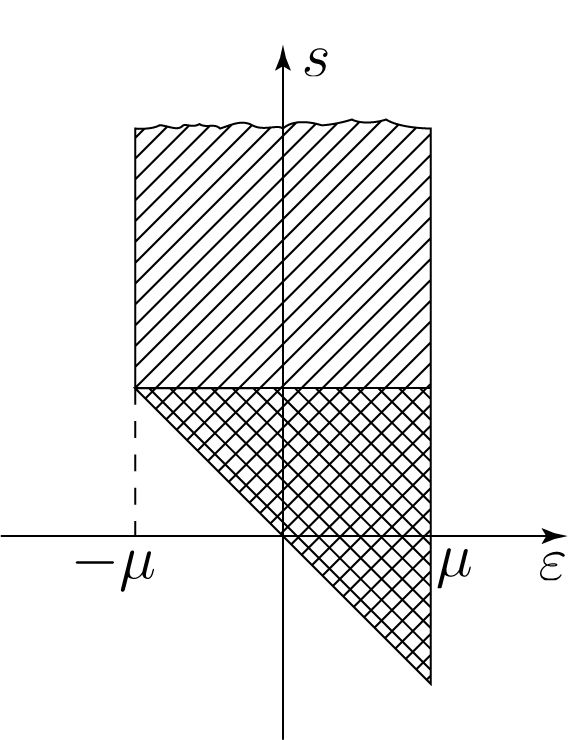}
 \caption{Integration region determining the particle density.}
\label{int1}
\end{figure}

 The summation can be changed by integration due to the smoothing of LLs. We introduce the variable $E = \sqrt{v^2p_z^2+n\Omega^2}$ and change summation over $n$ by integration over $dn$ also switching to polar coordinates where $\sqrt{n}\Omega = E\sin\theta$.
\begin{gather}
    \int\limits_{-\infty}^\infty dp_z \Omega^2\sum\limits_n \ \underset{W\gg\Omega}{\longrightarrow}\ \int\limits_0^\pi\sin\theta d\theta\int\limits_{0}^\infty E^2dE.
\end{gather}
Next, we change integration variable to $s=E-\ve$. The new integration region is depicted in Fig.~\ref{int1}.
Then we have
\begin{gather}
\label{app_density_strong_dis}
    n =\frac{1}{2\pi v^2\sqrt{2\pi}W} \int\limits_{-\mu}^\mu d\ve\int\limits_{-\ve}^\infty ds(s+\ve)^2 e^{-s^2/2W^2}.
\end{gather}
Changing the integration order we obtain
\begin{gather}
    n= \frac{\mu}{6\pi v^3}(\mu^2+3W^2).
\end{gather}
As a result, one needs to distinguish between two limits $\mu\gg W$ and $\mu\ll W$. The case $\mu\ll W$ is inconsistent with the electroneutrality condition. We have $\mu\sim T_{\rm imp}^3/W^2$.
Then, from the formula for the Debye length~\eqref{DebyeLength} we obtain $\kappa\sim \sqrt{\alpha }W/v \Rightarrow W \sim \sqrt{\alpha} T_{\rm imp}$ and $\mu\sim \alpha^{-3/2}W\gg W$ (contradiction).

Thus, for $W\gg\Omega$ only $\mu\gg W$ case is viable, and we have the following estimate
\begin{gather}
\label{doping_strong1}
  \mu\sim T_{\rm imp},\ \ W\sim\alpha^{3/4} T_{\rm imp}  \gg \Omega.
\end{gather}


\subsection{Weak disorder $W\ll\Omega$}
In this case the calculation is easily done for arbitrary  $\mu$. We can compute integrals in~\eqref{number} approximately using the strongly peaked nature of the exponential functions at weak disorder
\begin{gather}
    n = \frac{\Omega^2}{2\pi^2v^2}\sum\limits_{n=0}^{[\mu^2/\Omega^2]}\sqrt{\mu^2-n\Omega^2}
\end{gather}
In particular, if $\mu<\Omega$, we have the contribution to the particle density from the zeroth LL only
\begin{gather}
\label{app_density_weak_low}
    n = \frac{\mu\Omega^2}{2\pi^2v^2}.
\end{gather}
Therefore, condition $\mu< \Omega$ entails $T_{\rm imp}\lesssim\Omega$.

If $\mu\gg\Omega$ then we have
\begin{gather}
\label{app_density__weak_high}
    n = \frac{\mu^3}{3\pi^2v^2}
\end{gather}
As a result, we obtain the inverse Debye length in the form
\begin{gather}
\label{kappa_small}
    \kappa  = \sqrt{\alpha}\mu/v.
\end{gather}
Due to electroneutrality condition, we have $\mu\sim T_{\rm imp}$ and
\begin{gather}
    W\sim T_{\rm imp}\alpha^{3/4},
\end{gather}
as for the case of a strong disorder.


\begin{widetext}
\section{Perturbation theory for weak and average disorder $W\ll \Omega$}
\label{app2}
\subsection{First order corrections to the Green's functions}

To shed some light on the structure of perturbation series in the limit of weak and average disorder $W\ll\Omega$, let us analyze the perturbative expression for the imaginary part of the Green's function $n$th component in the first-order of the disorder strength $W$ (Fig.~\ref{abrikosov0}(a))
\begin{gather}
\label{green-corr}
    \begin{split}
        {\rm Im}G^{11}_n(\ve,\mathbf{p}_n)&\propto    \alpha^{(n)}_{0}+\sum\limits_{m\neq 0}\alpha^{(n)}_m\theta(\ve^2-m\Omega^2) ,\\
        \alpha_0^{(n)} &= \frac{W^2v\kappa}{\Omega^2 2^n n}
        \Big(E_n[a](n + a)e^a-1\Big),\quad
        a = l_H^2(\kappa^2+(\ve-p_z)^2) .
    \end{split}
\end{gather}

Here, $E_n(z) = \int_1^\infty e^{-zt}dt/t^n$ is the exponential integral.

We see that the mixing of the $n$th component of the Green's function with other LLs (of the order $m\neq0$) leads to $\theta$-function type terms in the imaginary part of the Green's function. It means that these terms do not contribute to conductivity in the limit of the small chemical potential $\mu < \Omega$ and zero temperature $T\rightarrow 0$.
The mixing with the zeroth LL term $\alpha_0^{(n)}$ lacks the $\theta$-function and can be significant.

In Eq.~\eqref{green-corr}, the term with $\ve-p_z$ is of the order of $W$. Therefore, for the average disorder strength
$ \kappa v \ll W\ll \Omega$, we have $a\approx l_H^2\kappa^2\ll 1$ and
\begin{gather}
\label{higherLL}
\begin{split}
    &{\rm Im} G_n^{11}(\ve,\mathbf{p}_n) \approx \frac{W^2 v\kappa}{\Omega^4 2^n n}\frac{1-\delta_{n,1}}{n-1}+\delta_{n,1}\ln\frac{1}{l_H\max\{\kappa, W\}},\ \ \frac{W}{\Omega}\ll 1 .
\end{split}
  \end{gather}
Here, we need to point out the following.  Equations~\eqref{green-corr} and \eqref{higherLL} tell us that the non-exponentially suppressed contribution to conductivity from higher LLs is possible in the the second order of perturbation theory in the disorder strength (see Fig.~\ref{diag_loop}).

\begin{figure}[h]
\centering
 \includegraphics[width=0.35\textwidth]{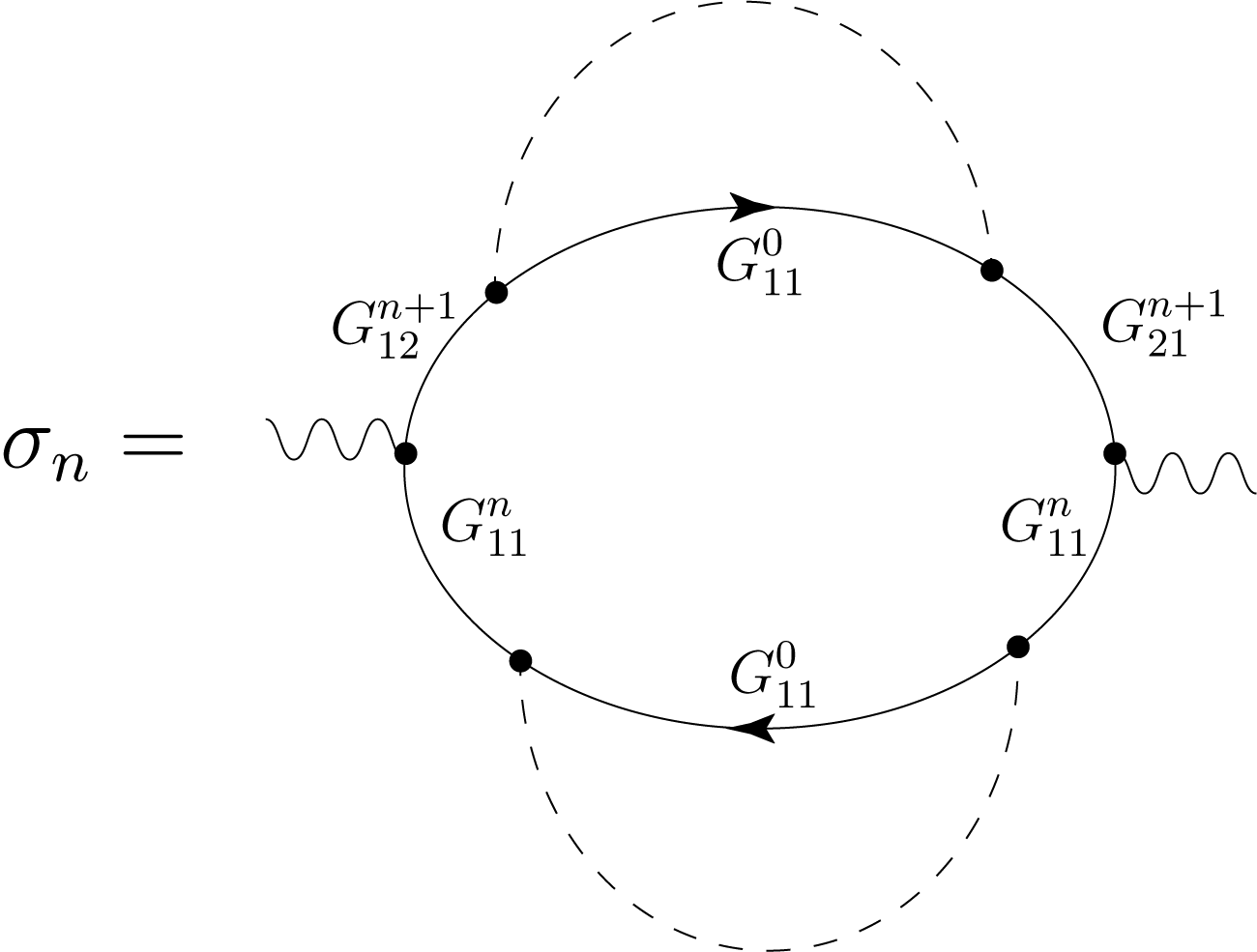}
 \caption{Second order (in the disorder strength) contribution to the conductivity.}
\label{diag_loop}
\end{figure}

On the other hand, the contribution from the zeroth LL is significant already in the first order of perturbation theory in the disorder strength (see Fig.~\ref{abrikosov0}). However, the contribution from higher LLs involves a summation over all levels with $n>1$. Therefore, an additional estimate is needed to understand whether we can discard higher LLS. We argue below in Section~\ref{higherLL} that higher LLs' contribution is indeed parametrically smaller and can be omitted.

\subsection{Contribution to conductivity from higher Landau levels}
\label{higherLL}
Using \eqref{higherLL} and the corresponding diagram (see Fig.~\ref{diag_loop}) we obtain
\begin{gather}
    \sigma_n(p_z) = \Big(\frac{W^2\kappa}{\Omega^2}\Big)^2\frac{v^2p_z^2}{\ve^4_n}\frac{1}{n(n-1)}\frac{(n+1)\Omega^2}{\ve_{n+1}^4}\frac{1}{n(n+1)},\quad n>1.
\end{gather}
Here we already set to zero the energy entering the Green's functions due to the strongly peaked Fermi function at $T\rightarrow 0$.
Their contribution to the conductivity then reads
\begin{gather}
    \sigma_{\rm high\ LLs}\propto \sum\limits_{n>1}\sigma_n(p_z)\frac{dp_z}{2\pi}\sim\frac{W^4\kappa^2}{v\Omega^7}
\end{gather}
On the other hand, the contribution from the zeroth LL reads (see Eq.~\eqref{higherLL}):
\begin{gather}
    \sigma_0 \sim\frac{W^2\kappa}{\Omega^4}\ln\frac{\Omega}{W}.
\end{gather}
As a result we see, that the contribution of the higher LLs is suppressed in the limit of the average disorder strength $\kappa v\ll W \ll \Omega$
\begin{gather}
    \sigma_{\rm high\ LLs} = \sigma_0\frac{W^2\kappa v}{\Omega^3}\Big[\ln\frac{\Omega}{W}\Big]^{-1}.
\end{gather}
Therefore, the contribution from high LLs is indeed suppressed.

\subsection{Classification of diagrams}
\subsubsection{Zeroth Landau level}
We are analyzing the expression for the loop diagram presented in Fig.~\ref{cond2}.  As was discussed above the only non-exponentially suppressed contributions are the ones, which include the Green's function components associated with the zeroth LL.
This means for the upper Green's function in the loop in Fig.~\ref{abrikosov0}(b) that it should facilitate \textit{the transition from the first LL to the zeroth one and back}.

First of all, let us convince ourselves that at each perturbation order it is possible to extract the class of diagrams, which provides the leading contribution.
Let us compare analytical expressions for two second order diagrams (see. Fig.~\ref{diag2}).

Let us introduce the notation
\begin{gather}
    K(q_1,q_2,q_3)= \chi_{1}(x_0)\chi_{0}(x_{q_1})
        \chi_{0}(x_{1,q_1})\chi_{0}(x_{1,q_2})
        \chi_{0}(x_{2,q_2})\chi_{0}(x_{2,q_3})
        \chi_{0}(x^\prime_{q_3})\chi_{1}(x_4).
\end{gather}

\begin{figure}[h]
\centering
 \includegraphics[width=0.65\textwidth]{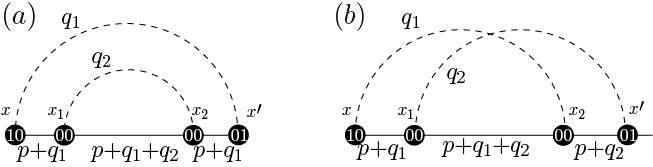}
 \caption{Two possible contribution in the second order of perturbation theory.}
\label{diag2}
\end{figure}

Here, we remind the reader the notation $x_{q} \equiv x-q_y l_H^2$.
The expressions for the diagrams in Fig.~\ref{diag2} have the form:
\begin{gather}
\label{sigma2}
    \begin{split}
    &\Sigma^{(R,2)}_a = \int
    e^{iq_{1,x}(x_0-x_4)}e^{iq_{2,x}(x_1-x_2)}K(q_1,q_1+q_2,q_1)
       [d^{4}x][G^R_{p_z+q_{1,z}}]^2G^R_{p_z+q_{1,z}+q_{2,z}}g(\mathbf{q_1})g(\mathbf{q_2})d\Gamma_{q_1}d\Gamma_{q_2},\\
   &\Sigma^{(R,2)}_b = \int
    e^{iq_{1,x}(x_0-x_2)}e^{iq_{2,x}(x_1-x_4)}K(q_1,q_1+q_2,q_2)
       [d^{4}x]G^R_{p_z+q_{1,z}}G^R_{p_z+q_{2,z}}G^R_{p_z+q_{1,z}+q_{2,z}}g(\mathbf{q_1})g(\mathbf{q_2})d\Gamma_1d\Gamma_2.
    \end{split}
\end{gather}
Let us compute the integrals over the coordinates entering the first and second diagrams
\begin{gather}
\label{sigmaint}
    \begin{split}
    & I_{a}(\mathbf{q}_{\bot,1}, \mathbf{q}_{\bot,2}) = \int
    e^{iq_{1,x}(x_0-x_4)}e^{iq_{2,x}(x_1-x_2)}K(q_1,q_1+q_2,q_1)
       [d^{4}x] = \frac{1}{8} e^{-\frac{1}{2}l_H^2(q_{1,\bot}^2+q_{2,\bot}^2)}l_H^2q_{1,\bot}^2,\\
   & I_{b}(\mathbf{q}_{\bot,1}, \mathbf{q}_{\bot,2}) = \int
    e^{iq_{1,x}(x_0-x_2)}e^{iq_{2,x}(x_1-x_4)}K(q_1,q_1+q_2,q_2)
       [d^{4}x] = \frac{1}{8} l_H^2 q_{1\bot}q_{2\bot} e^{-l_H^2(q_{1,\bot}^2+q_{2,\bot}^2)}
       e^{iq_{1,\bot}q_{2,\bot}l_H^2\sin(\theta_1-\theta_2)-i(\theta_1-\theta_2)}.
    \end{split}
\end{gather}
Here, $q_{\bot,i} \equiv |\mathbf{q}_{\bot,i}|$ and $\theta$ is the polar angle of the direction of $\mathbf{q}_\bot$.
The potential does not depend on the direction of $\theta$. Therefore, we can perform integration over $\theta_{1,2}$ in the expression for self-energies~\eqref{sigma2} (deciphering $d\Gamma_q$ as $d\theta q_{\bot}dq_{\bot}dq_z (2\pi)^{-3}$). We obtain
\begin{gather}
\label{sigmaint2}
    \begin{split}
         \int I_{\sigma}(\mathbf{q}_{\bot,1}, \mathbf{q}_{\bot,2})\frac{d\theta_1d\theta_2}{(2\pi)^2} = \frac{1}{8} e^{-\frac{1}{2}l_H^2(q_{1,\bot}^2+q_{2,\bot}^2)}
         \begin{cases}
            l_H^2q_{1,\bot}^2,\ &\ \sigma =a,\\
            l_H^2q_{1\bot}q_{2\bot} J_{1}(l_H^2q_{1\bot}q_{2\bot}),\ &\ \sigma =b.
         \end{cases}
    \end{split}
\end{gather}
Here $J_1(z)$ is the Bessel function.
The integrals $I_{a,b}(q_{1,\bot}, q_{2,\bot})$ have different dependence on the small parameter of the system, $\kappa l_H\sim\sqrt{\alpha}$ when being integrated over momenta moduli $q_{1,\bot}$ and $q_{2, \bot}$ with potential correlation function. Using the dimensional variable $s = q_{\bot}l_H$, we rewrite these integrals as
\begin{gather}
\label{sigmaint3}
    \begin{split}
         &\int\frac{dq_{1,z}dq_{2,z}}{(2\pi)^2}\int I_{\sigma}(\mathbf{q}_{\bot,1}, \mathbf{q}_{\bot,2})\frac{d\theta_1d\theta_2}{(2\pi)^2}
         \frac{q_{1,\bot}dq_{1,\bot}q_{2,\bot}dq_{2,\bot}}{(2\pi)^2}
         \\
         &= \frac{ l_H^2}{32\pi^2}\int\frac{dq_{1,z}dq_{2,z}}{(2\pi)^2}\int\limits_{0}^\infty \frac{e^{-\frac{1}{2}(s_1^2+s_2^2)}}{(s_1^2+\Delta_1^2)^2
            (s_2^2+\Delta_2^2)^2}
         \begin{cases}
            s_1^3ds_1s_2ds_2,\ &\ \sigma =a,\\
            s_1^2 s_2^2 J_{1}(s_1s_2)ds_1ds_2,\ &\ \sigma =b,
         \end{cases}\\
         \Delta_{i}^2 &= l_H^2[q_{i, z}^2+\kappa^2]
    \end{split}
\end{gather}
The value of momentum $q_z$ is going to be in a range $q_z \in [0,\sim W)$. In the limit $l_H\kappa \rightarrow 0$ integrals show different infrared behavior. The integral containing $I_a$ contains infrared divergence while the one with $I_b$ remains convergent
\begin{gather}
  \begin{split}
    &\int I_a\sim\int_{\kappa l_H}\frac{ds_1s_1^3 dq_{1,z}}{(s_1^2+q_z^2)^2} \int_{\kappa l_H}\frac{s_2ds_2}{(s_2^2+q_{2,z}^2)^2}\sim\frac{1}{\kappa l_H}, \\
    &\int I_b\sim\int_{\kappa l_H}\frac{s_1^3 ds_1}{(s_1^2+q_{1,z}^2)^2} \int_{\kappa l_H}\frac{ds_2 s_2^3}{(s_2^2+q_{2,z}^2}\sim 1,
  \end{split}
\end{gather}

\begin{gather}
    \Sigma_b\sim (\kappa l_H) \Sigma_{a}\ll \Sigma_{a}.
\end{gather}
Therefore, we realize that this type of self-energy diagrams with crossing disorder lines is always suppressed as compared to the ones without crossings.

\subsubsection{Contribution to the self-energy from higher Landau levels}
Identical calculation shows that expressions~\eqref{sigmaint}
retain its exponential form and preexponential momentum dependence for any pair $(n, n+1)$ of LLs. This leads to the conclusion that the same classification of diagrams holds for any pair $(n,\ n+1)$ of LLs.

We need to check that the contribution from the diagrams $(n, n+m)$ is suppressed (see Fig.~\ref{numerics}(a)).
To this end, we compute the matrix element defining the self energy
\begin{gather}
    f(n,n+m) = \int\frac{dq_x dq_y}{(2\pi)^2}\int dx dx^\prime \chi_n(x_{p_y})\chi_{n+m}(x_{p_y}-q_y l_H^2)
    \chi_{n+m}(x^\prime_{p_y}-q_y l_H^2)\chi_n(x^\prime_{p_y})g(\mathbf{q})e^{iq_x(x-x^\prime)}.
\end{gather}

\begin{figure}[t]
\centering
 \includegraphics[width=1.0\textwidth]{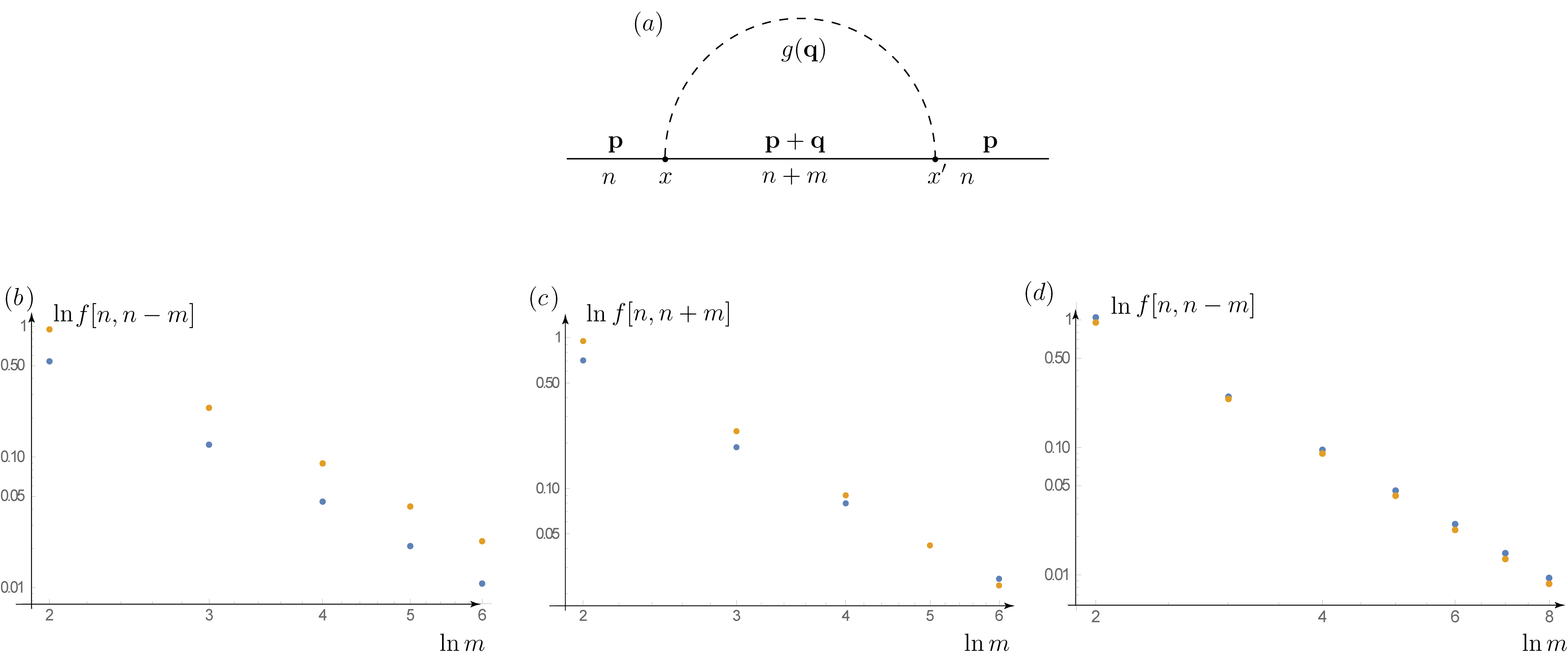}
 \caption{Different matrix elements $f(n, n+m)$ determining the self-energy:  (a) The diagram for the self energy;   (b) $n=7$,  the mixing comes from lower LLs up to $n=1$, (c) $n=7$, the mixing comes from upper LLs up to $n=13$; (d) $n=13$, the mixing comes from lower LLs up to $n=5$.
 The blue dots correspond to exact values, while orange dots represent the fitting by function $f(m) = x^{-3.4}$.
 }
\label{numerics}
\end{figure}
The numerical analysis shows that the matrix elements decay approximately as $\sim m^{-3.4}$ (see Fig.~\ref{numerics}(b-d)).
On the other hand, the leading term yields
\begin{gather}
    \label{main1}
    f(n, n+1) \approx\ln\frac{\Omega^2}{W^2}.
\end{gather}

Therefore, the summation of the matrix elements $f(n, n+m)$ over $|m|>1$ gives the sub-log behavior comparing to the principal one~\eqref{main1}.

\subsection{Vertex diagrams}
Incredibly, in the ultraquantum limit, all vertex corrections disappear. This happens due to  orthogonality of LLs wave functions, which are interleaved by diagrammatic loop structure into very specific contractions. Instead of giving general arguments, we present some very characteristic examples of diagrams which show the general idea of the mechanism (see Fig.~\ref{diag2}).
\begin{figure}[H]
\centering
 \includegraphics[width=0.4\textwidth]{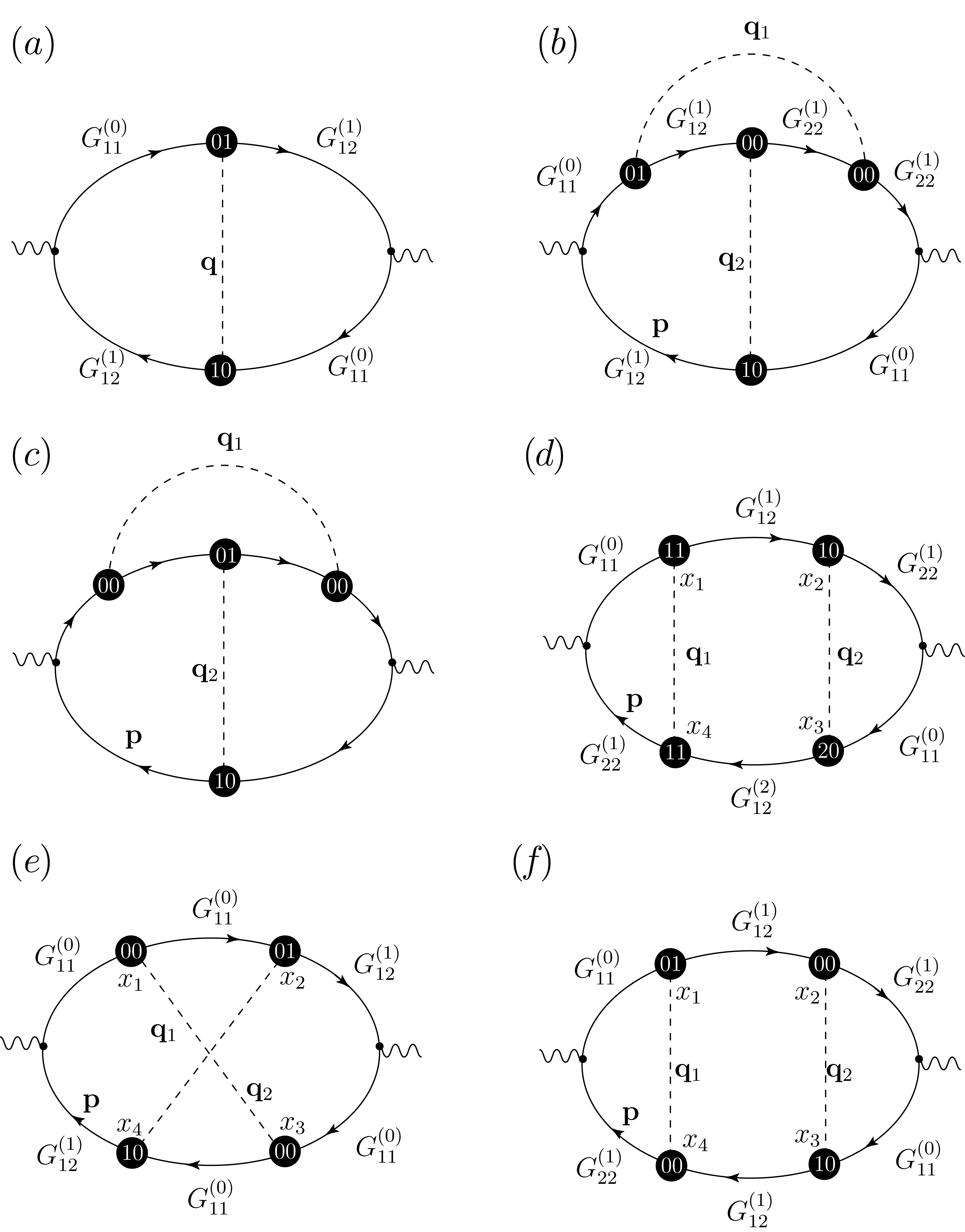}
 \caption{Typical vertex diagrams in the first and second order of the perturbation theory.}
\label{diag2}
\end{figure}
The estimate of each diagram in Fig.~\ref{diag2} is as follows
\begin{gather}
    \begin{split}
            &(a)\sim e^{2i\theta},\quad (b)\sim e^{iq_{1,\bot}q_{2,\bot}l_H^2\sin(\theta_1-\theta_2)+2i\theta_1+i\theta_2},\quad (c)\sim e^{iq_{1,\bot}q_{2,\bot}l_H^2\sin(\theta_1-\theta_2)+2i\theta_1+i\theta_2},\\
    &(d)\sim e^{2i\theta_2},\quad (e)\sim e^{iq_{1,\bot}q_{2,\bot}l_H^2\sin(\theta_1-\theta_2)+2i\theta_2},\quad (f)\sim e^{i\theta_1+i\theta_2}.
    \end{split}
\end{gather}
Here, $\theta,\ \theta_1,\ \theta_2$ are the guiding angles of momenta $\mathbf{q}_\bot$. We see that every contribution disappears after integration over guiding angles. The pattern persists in all higher orders of perturbation theory.

\subsection{Computation of the conductivity at average disorder $\kappa v\gg W\ll\Omega$ and small doping $\mu < \Omega$}
The exact Green's functions now are to be computed formally in all orders of perturbation theory. This problem seems intractable. However, the result should be the analytic function of the small parameter of our system: $\kappa l_H\sim\sqrt{\alpha}$. Therefore, we expect $G^{R}=G^R|_{\alpha = 0} +\mathcal{O}(\sqrt{\alpha})$. The expression for $\alpha =0$ corresponds to the treatment by the Keldysh model.

Therefore, in the lowest order in $\kappa l_H\ll1$ and in all orders of perturbation theory in disorder strength $W/\kappa$, the conductivity is given by skeletal diagram (Fig.~\ref{diag2}).

Therefore, the Kubo loop in the leading order in $\kappa l_H$ reads
\begin{gather}
    {\rm Loop} = \int\frac{dp_z}{2\pi}{\rm Im}\big\{\mathbf{G}^{(0)}_{11}(p_z)\big\}
    \big[{\rm Re}\mathbf{G}^{(1)}_{21}(p_z)\big]^2
    \int\frac{dq_z}{2\pi}{\rm Im}\big\{\mathbf{G}^{(0)}_{11}(p_z+q_z)\big\}g_{\rm eff}(q_z),\\
    g_{\rm eff}(q_z) = W^2l_H^2\kappa\int\limits_{0}^\infty e^{-l_H^2q_\bot^2}\frac{1}{(q_\bot^2+q_z^2+\kappa)^2}\frac{q_\bot^3 dq_\bot}{16\pi}\approx
    \frac{W^2l_H^2\kappa}{16\pi}\ln\frac{\Omega}{\max\{W, p_z\}}.
\end{gather}
Next, we compute the diagrams in the framework of the Keldysh model
\begin{gather}
\begin{split}
    {\rm Re}\mathbf{G}^{(1)}_{21} &= \frac{\sqrt{2}\Omega}{\sqrt{\Omega^2+v^2 p_z^2} W}D_+\Big(\frac{\sqrt{\Omega^2+v^2 p_z^2}}{\sqrt{2}W}\Big),\quad D_+(x) = e^{-x^2}\int_0^x e^{t^2}\,dt,\\
    {\rm Im} \mathbf{G}^{(0)}_{11}(p_z) &=-\sqrt{\frac{\pi}{2}}\frac{1}{W}e^{-\frac{p_z^2}{2W^2}}.
\end{split}
\end{gather}
To compute the skeletal self-energy we evaluate the integral
\begin{gather}
    \mathbf{\Sigma}(p_z) = \int{\rm Im} \mathbf{G}^{(0)}_{11}(p_z+q_z)g_{\rm eff}(q_z)\frac{dq_z}{2\pi}.\quad
\end{gather}
In the  limit $W\gg\kappa v$,  we can discard $\kappa$ in the denominator of the expression defining $g_{\rm eff}$. Then, the integral defining $\Sigma$ can be expressed as
\begin{gather}
    \mathbf{\Sigma}(p_z) \approx \frac{W^2\kappa l_H^2}{16\pi v}
    \int\limits_{-\pi/2}^{\pi/2}e^{-pz^2/2W^2\cos^2\vf}\frac{d\vf\cos^3\vf}{(\cos^2\vf+\frac{4W^2}{\Omega^2}\sin^2\vf)^2},\quad \kappa l_H\ll1.
\end{gather}
For $\Sigma$ in the limiting case  $\kappa v\ll W\ll\Omega$, we obtain
\begin{gather}
    \Sigma(p_z)\approx \frac{W^2\kappa l_H^2}{16\pi v}
        \ln\frac{\pi^2\Omega^2}{16\max\{p_z^2, W^2\} },\ \ v\kappa\ll W\ll\Omega.
\end{gather}
The first asymptotics is obtained with the log accuracy.
Now we can compute the Kubo loop determining the conductivity as follows
\begin{gather}
\label{loop2}
    {\rm Loop} =\frac{\sqrt{2\pi}}{v}\frac{\Omega^2}{W^4} \int\limits_{-\infty}^\infty\frac{ds}{2\pi} e^{-s^2/2}\Sigma(s W)\frac{1}{s^2+\Big(\frac{\Omega}{W}\big)^2}D^2_+\left[\sqrt{\frac{1}{2}\left(s^2+\frac{\Omega^2}{W^2}\right)}\right].
\end{gather}
Finally, using the asymptotics for the $D_+(x)\approx(2x)^{-1},\ x\rightarrow+\infty$, we compute the Gaussian integral entering~\eqref{loop2} and obtain Eq.~\eqref{cond_fin}.

\subsection{Computation of the conductivity, high doping $\mu \gg \Omega$}
The computation proceeds essentially along the same lines as for the low doping.
There is a small difference coming from the effective Coulomb interaction. The latter one  with the log accuracy reads
\begin{gather}
    g_{\rm eff, n} \approx W^2l_H^2\kappa\frac{n+1}{2}\int_0^{\sim l_H^{-1}}\frac{1}{(q_\bot^2+q_z^2+\kappa^2)}\frac{q_\bot^3 dq_\bot}{16\pi}\approx
    W^2l_H^2\kappa\frac{n+1}{32\pi}\ln\frac{\Omega}{W}.
\end{gather}
Here, we take into account the mixing of $n$ and $n+1$ LLs only. Recall that the contribution from the mixing of $n$ and $n+m$ LLs is proportional to the matrix element, which approximately decays as $\sim1/m^{3.4}$.

Then, the self-energy reads in the leading log accuracy yields:
\begin{gather}
    \Sigma_n(p_z)\approx \frac{W^2\kappa l_H^2(n+1)}{32\pi v}\ln\frac{\Omega}{\max\{p_z,W\}}\theta(\mu-\sqrt{n}\Omega).
\end{gather}
Finally, the conductivity takes the form
\begin{gather}
    \sigma_{xx} = \frac{W^2\kappa l_H^2}{32\pi v}\ln\frac{\Omega}{W}\Omega^2\sum\limits_{n=0}^{\frac{\mu^2}{\Omega^2}}\int\limits_{-\infty}^\infty\frac{1}{2\sqrt{2\pi}W}e^{-[\ve_n(p_z)-\mu]^2/2W^2}\frac{(n+1)^2}{\Omega^2}\frac{dp_z}{2\pi}.
\end{gather}
Integrating over momentum $p_z$ (by expanding it in the vicinity of the root of the equation $\mu^2 = n\Omega^2+v^2p_z^2$), we arrive at the approximate formula~\eqref{cond_high_mom} (here, we assume that $\mu^2/\Omega^2$ is not an integer).

If  $\mu^2/\Omega^2\in \mathbb{N}$, the term with $n = \frac{\mu^2}{\Omega^2}$ yields different contribution. Expanding near the maximum the exponent of the exponential, we obtain
\begin{gather}
    \sigma_{xx} = \frac{e^2W^2\kappa\mu}{16\pi^2\Omega^2}\ln\frac{\Omega^2}
    {W^2}\bigg[\sum\limits_{n=1}^{\frac{\mu^2}{W^2}-1}
    \frac{\mu(n+1)^2}{\sqrt{\mu^2-n\Omega^2}}+ 2^{3/4}\sqrt{\frac{\mu}{W}}\Gamma\Big(\frac{5}{4}
    \Big)\Big(\frac{\mu^2}{\Omega^2}+1\Big)^2\bigg],
\end{gather}
which yields the formula ~\eqref{cond_high_mom} in the main part of the paper.

\section{Hall conductivity, strong disorder, $ W\gg\Omega$}
The computation of the Hall conductivity at strong disorder differs slightly from that at the weak disorder one due to the largeness of the chemical potential. The anomalous part can be writtne as
\begin{gather}
    \sigma_{xy}^{II}=ec\frac{\partial}{\partial H}n(\mu) \equiv 2ecv^2\frac{\de}{\de \Omega^2}n(\mu),
\end{gather}
and after simple algebra we obtain
\begin{gather}
    \sigma_{xy}^{II}=\frac{e^2\Omega^2}{4\pi}\int\limits_{-\infty}^{\infty}\frac{dp_z}{2\pi}\frac{n}{2\ve_n}\frac{1}{\sqrt{2\pi}W}\Big(e^{-\frac{(\ve_n-\mu)^2}{2W^2}}-e^{-\frac{(\ve_n+\mu)^2}{2W^2}}\Big) =\frac{e^2\mu^3}{6\pi^2v\Omega^2}.
\end{gather}
Here, we switched from the summation to the integration formula, exploiting the smoothness of the summand due to the condition $W\gg\Omega$. This result should be compared with the one for the weak disorder presented in Ref.~\onlinecite{KlierPRB2017}.

Here, we also exploited the condition $\mu\gg W$.
Now, in the same manner, we compute the normal part.
The straghtforward computation  of~\eqref{sigma_xy0}  reveals that
\begin{gather}
    \sigma_{xy}^{I}=\sigma_{xy}^{II}\ \Rightarrow \sigma_{xy} = \frac{e^2\mu^3}{3\pi^2\Omega^2}
\end{gather}

\vskip10cm
\end{widetext}

\bibliographystyle{apsrevlong_no_issn_url}

\bibliography{quantumMR}

\end{document}